\documentclass[11pt]{article} 
\usepackage{amsmath} 
\usepackage{amssymb}
\usepackage{graphicx} 
\usepackage{float} 
\usepackage{color} 
\usepackage{array,  graphicx} 
  
\numberwithin{equation}{section} 

\def\lowcomma{_{\textstyle,}}

\title{Estimation and Confidence Intervals for Mutual Information: Issues in Convergence for Non-Normal Distributions }
\author {by Theo Grigorenko and Leo Grigorenko}
\begin{document} 
\maketitle

\begin{quotation} 
\noindent {\it Abstract:}

By employing various empirical estimators for the Mutual Information (MI) measure, we calculate and compare the estimates and their confidence intervals for both normal and non-normal bivariate data samples. We find that certain nonlinear invertible transformations of the random variables can significantly affect both the estimated MI value and the precision and asymptotic behavior of its confidence intervals. Generally, for non-normal samples, the confidence intervals are larger than those for normal samples, and the convergence of the confidence intervals is slower even as the data sample size increases. In some cases, due to strong biases, the estimated confidence interval may not contain the true value at all. We discuss various strategies to improve the precision of the estimated Mutual Information.

\vspace{9pt} 

\end{quotation}

\section {Introduction}

For bivariate data samples the sample correlation coefficient is a statistic of the true (population) correlation coefficient. Since Fisher \cite{fisher1,fisher2} it is known how to estimate the confidence intervals for a sample correlation coefficient under normality assumptions. When the sampling distribution for the sample correlation coefficients is not normal, one can achieve approximate normality (especially given jointly normal bivariate data) through the Fisher Z-transformation \cite{fisher1,fisher2}, thus yielding approximately $N^{-1/2}$ convergence of the confidence intervals, where $N$ is the length of the data. 

The Fisher transformation is routinely applied in many cases of non-normal bivariate random variables. However, it has been shown \cite{lognormal} that for lognormal bivariate data the convergence of the confidence intervals is much slower than $N^{-1/ }$ and the bias in estimating $\rho$ can be very large for $N<10^5$. This result suggests that for strongly non-normal data, the estimate can be unreliable even for a quite large length ($N \approx10^5$) of the data. 
Nowadays the data analysis for numerous applications requires a more general association measure, which may capture potential non-linearities. The mutual information measure, originally introduced by the pioneer of information theory, Shannon \cite { shannon1, shannon2}, satisfies this criterion. 
Mutual information $MI (X;Y)$ is commonly defined as: 
\begin{equation}
\label{definition} 
MI (X;Y) = \mathbb{E}\left( \log \frac{f_{x,y}}{f_x f y} \right) = \sum _{x,y} f_{x,y} (\log f_{x,y} - \log f_x f_y), 
\end{equation} where $X$ and $Y$ are two random variables with probability density function $f_x$ and $f_y$, respectively, and joint probability density functions $f_{x,y} $, and $\mathbb{E}$ is the expectation operator over $X$ and $Y$. As the Kullback-Leibler divergence \cite{KL_divergence} between the joint density and the product of the individual densities, it is able to measure the most general association between two random variables \cite{MI_measure1, MI_measure2}. It is worth mentioning that in the case of binary discrete variables the mutual information can be expressed as a function of the correlation coefficient \cite{MI_bivariate}. 

MI has many interpretations such as the reduction of uncertainty in $Y$ after observing $X$, or the stored information in one variable also contained in the other. This expanded applicability of mutual information over the correlation coefficient allows for expanded use, from extracting the dependence between both numeric and symbolic sequences (e.g. symbolic dynamics) to the study of phase trajectories in chaos dynamics. Further uses involve MI applications in finance \cite{MI_finance, finance_networks}, as well as diverse applications in analyses of human electroencephalograms \cite{MI_eeg}, corticomuscular interactions \cite{neuro_physiology}, image registration \cite{image_registration}, hierarchical data clustering \cite{clustering}, biochemical signaling systems \cite{bio_signaling} and gene association networks \cite{gene_networks}.

For bivariate data samples from a general population, the mutual information calculated on each of these samples is, just like the sample correlation coefficient, a statistic or a point estimator of the true mutual information population measure \cite{MI_estimation1, MI_estimation2}. Naturally then, there should also be a way to devise the construction of confidence intervals for statistical inference of mutual information---akin to inference for linear correlation. Since MI is a more general measure, one would expect that the confidence intervals for a nonlinear measure may be larger than those for the Pearson coefficient, and that convergence and bias studies for the linear coefficient in the case of non-normal marginals in bivariate data \cite{lognormal} may be more pronounced for the MI measure. In this paper, we study the construction and behavior of the confidence intervals for the mutual information measure under different parametric assumptions and transformations; our findings show how the reliability of statistical inference for the MI measure can decrease when considering several estimators under various, largely non-normal, circumstances.

\section{Calculating Mutual Information \& Confidence Intervals: Methodology \& Results} 
For a normally distributed vector $(X,Y)$ with mean values $\mu_i$, standard deviations $\sigma_i$, $i=1,2$, and correlation coefficient $\rho$, the mutual information between $X$ and $Y$ can be calculated analytically: \begin{equation} 
\label{MI_normal} 
MI_{Gauss} (X;Y) = - \frac{1}{2} \log(1-\rho^2).
\end{equation}	

Though the Gaussian case is useful analytically, many real--world instances, such as in financial or natural phenomena, exhibit non-normal, heavy-tailed, and often power-law distributional forms.
In order to estimate the reliability of our MI estimates in cases like those, we run simulations using the Student-t bivariate distribution with relatively low degrees of freedom (e.g. $ \nu=3$). The simulated results are benchmarked against the known  analytical result \cite{student} :

\begin{eqnarray} 
\label{studentt} 
MI_{Student-t} (X; Y) = MI_{Gauss} (X;Y) +2 \log\left(\sqrt{\frac{\nu}{2\pi}} 
\hspace{1mm} B\left(\frac{\nu}{2} \lowcomma 
\frac{1}{2} \right) \right) - \nonumber\\\frac {2+\nu}{\nu}+\left (1+ \nu \right) \left[\psi\left( \frac{ \nu+ 1}{2} \right) -
\psi\left(\frac{\nu}{2} \right) \right],
\end{eqnarray} 
where $B(x,y)$ is the beta function and $\psi(x) $ is the digamma function. Note that only the first term in Eq.~(\ref{studentt}) survives as $\nu \to \infty$.

It is known that the mutual information measure is invariant under smooth, uniquely invertible maps $X'$ and $Y'$ \cite{Kraskov}: 
\begin{equation} 
\label{transformation}
	MI (X;Y) =MI (X';Y'). 
\end{equation} 
This allows us to test widely-used numerical algorithms to estimate MI for a range of non-normal random variables under non-linear transformations. In this work, we present the results for the case of cubic transformations $X'=X^3$, $Y'=Y^3$ of the variables.

We performed the MI calculations and estimation of the confidence intervals (CI) with the Kraskov-Stoegbauer-Grassberger (KSG) kNN estimator \cite{Kraskov}, using $k=4$ nearest neighbors as default. This algorithm was benchmarked against the simple plugin estimator based upon the direct implementation of Eq.~(\ref{definition}). We find that the KSG algorithm outperforms the plugin method which, for given precision, would require a much larger sample, since it is based on the empirical estimation of the histograms for $f_{x,y} $ as well for $f_x, f y$.

The calculations were performed for different sample lengths, varying from $N=100$ to $N=10^6$. The corresponding CI were estimated by repeating the calculations for specific sample size for $R=1000$ times, and calculating the $5\%$ sample quantile as the lower bound, and the $95\%$ sample quantile as the upper bound. The results were compared with the analytical formulas in Eqs.~(\ref{MI_normal}, \ref{studentt}).

\subsection { Normality \& Log-Normality} 
\begin{figure}
\centering 
\includegraphics [width=0.8\textwidth] {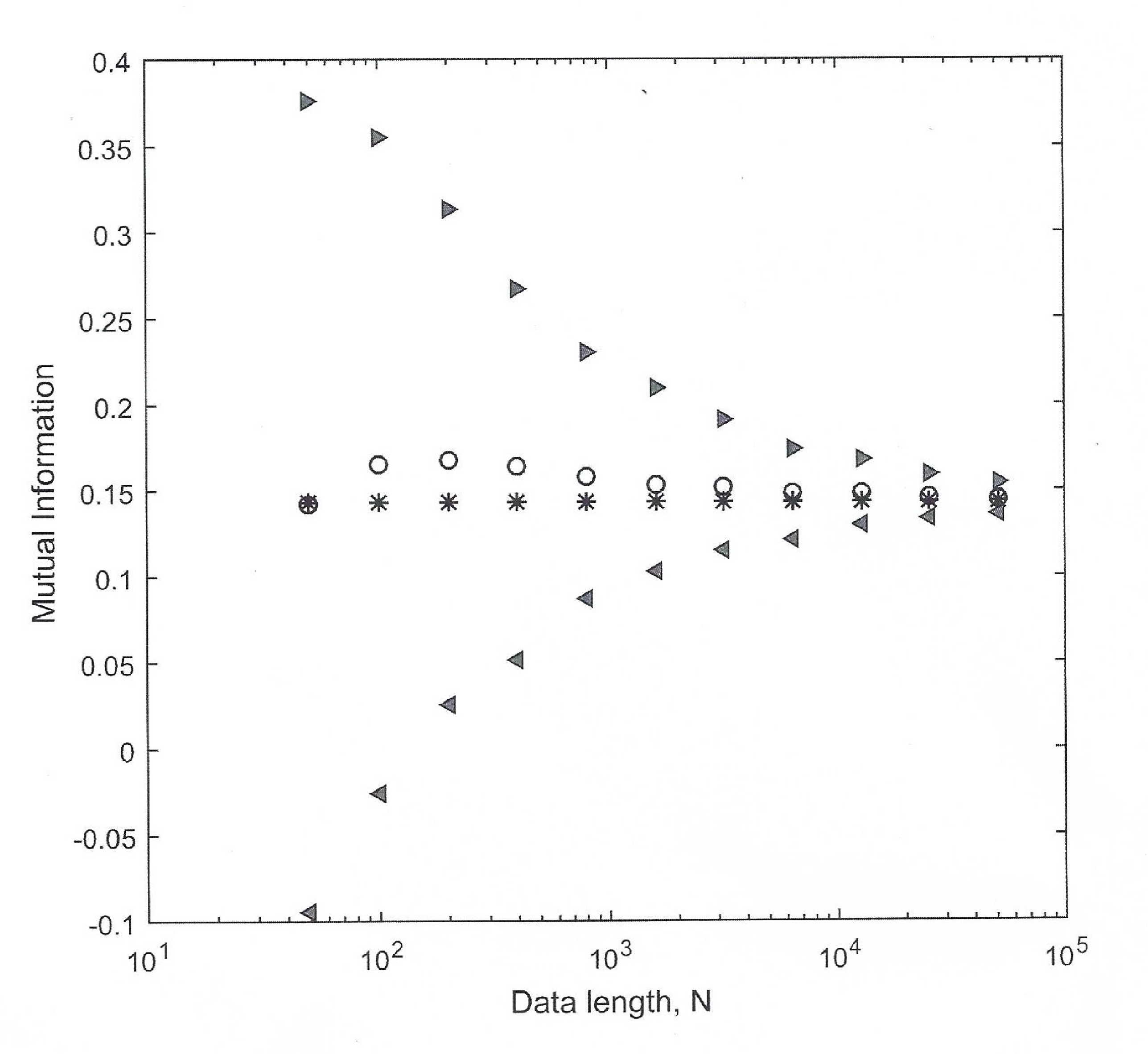}
\caption{$5\%$ and $95\%$ quantile confidence intervals ($``\bigtriangleup $"), the estimated mean MI value (``o"), 
and the analytical solution (``'*") as functions of the data length $N$. This is the case of normal bivariate data,
$\sigma_1=\sigma_2=1$, $\rho=0.5$. The numerical method used is the Kraskov-Stoegbauer-Grassberger estimator (see text). } \label{fig:image0} 
\end{figure}

We choose normal bivariate series with zero means $\mu_i=0$, $i= 1,2$, and equal standard deviations $\sigma_1 = \sigma_2 = 1$. The correlation coefficient was set to $\rho=0.5$. The obtained results are shown in Figure~\ref{fig:image0}. Note that for relatively small data sets ($50<N<300$) the KSG estimator may give a negative number for the non-negatively defined MI measure. Another observation is that even for a relatively large dataset lengths $10^4<N<10^5$ the confidence intervals are relatively big. However, the average of the $R=1,000$ ensemble estimates is close to the analytical value $MI=0.1438$. This suggests that a simple bootstrapping of the data can considerably improve the estimates of the mutual information measure. The results of calculations also show that the confidence intervals scale as $N^{-1/2}$ and there is no significant bias.

\begin{figure} [p] 
\centering 
\includegraphics [width=0.8\textwidth] {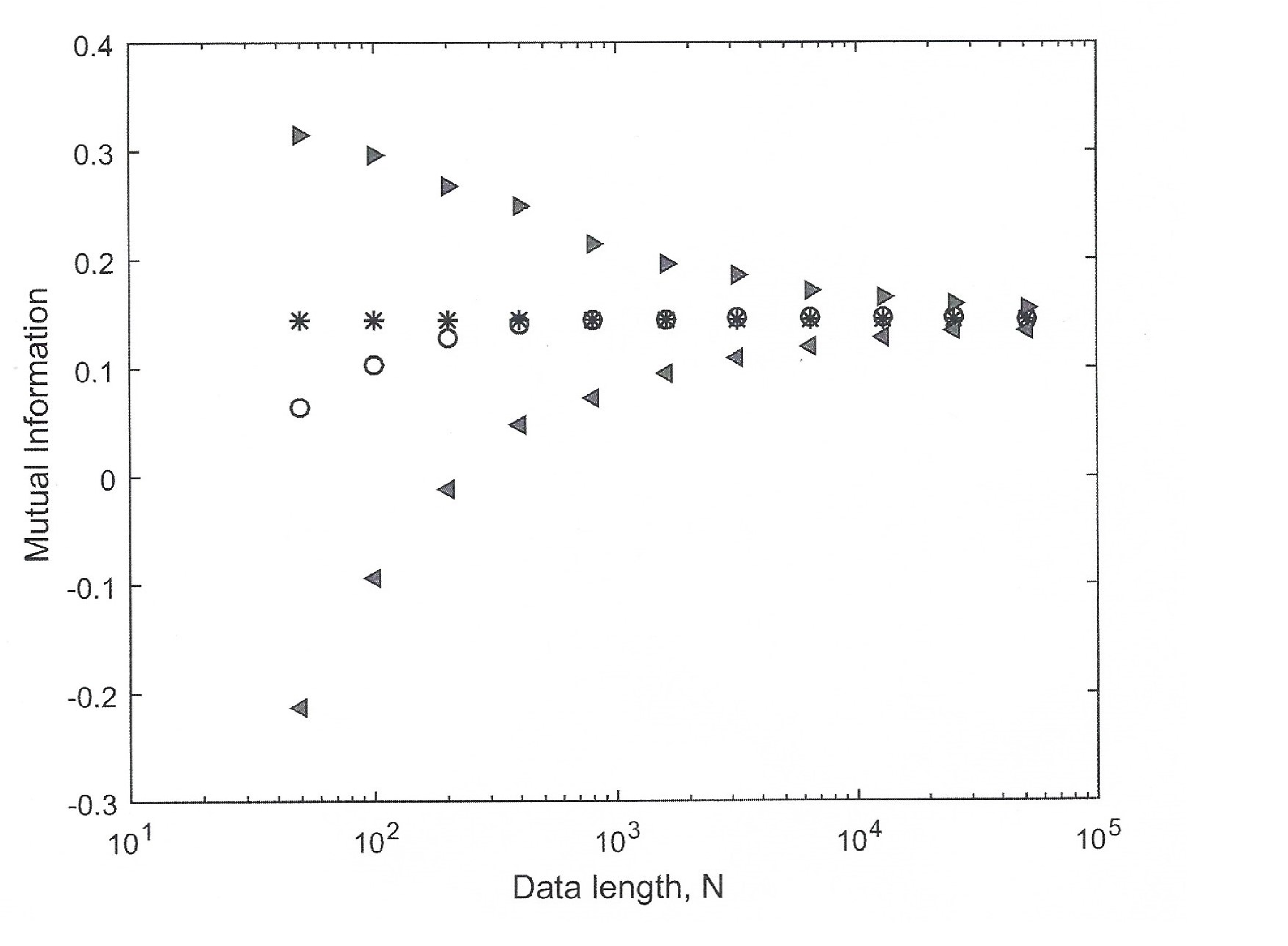} 
\caption{$5\%$ and $95\%$ quantile confidence intervals ($`` \bigtriangleup $"), the estimated mean MI value (
``o"), and the analytical solution (``*") as functions of the data length $N$. This is the case of lognormal bivariate data, 
$\sigma_1=\sigma_2= 1$, $\rho=0.5$. The numerical method used is the KSG estimator.} 
\label{fig:image1} 
\end{figure}

As our next step, we consider log-normally distributed vectors $(X, Y)$. Since the exponential transformation is a homeomorphism, the analytical MI value calculated for the corresponding normal distribution Eq.~(\ref{transformation}) remains unchanged. As in the previous case, we assume zero mean values $ \mu_1 = \mu_2=0$, the standard deviations $\sigma_1=\sigma_2=1$ and the correlation coefficient $\rho=0.5$. The results are shown in Figure~\ref{fig:image1}. Again, for relatively small data sets ($50<N<300$), the KSG estimator may produce a negative number, and the confidence intervals are relatively big still for $10^4 < N <10^5$. Similar to the normal case, the average of the $R=1,000$ ensemble estimates is close to the analytical value, the calculated confidence intervals Scale as $N^{-1/2}$ and there is no significant bias. We compared the performance of KSG estimator with the simple plugin method, and found that for the case of lognormal variables, the plugin method shows a significant bias, which is still present for data lengths $10^5<N<10^6$, which cannot be corrected by 
the standard Miller-Madow estimator \cite{MI_estimation1,MI_estimation2}.

\subsection{Non-Normality} 
\begin{figure} [p] 
\centering
\includegraphics [width=0.8 \textwidth] {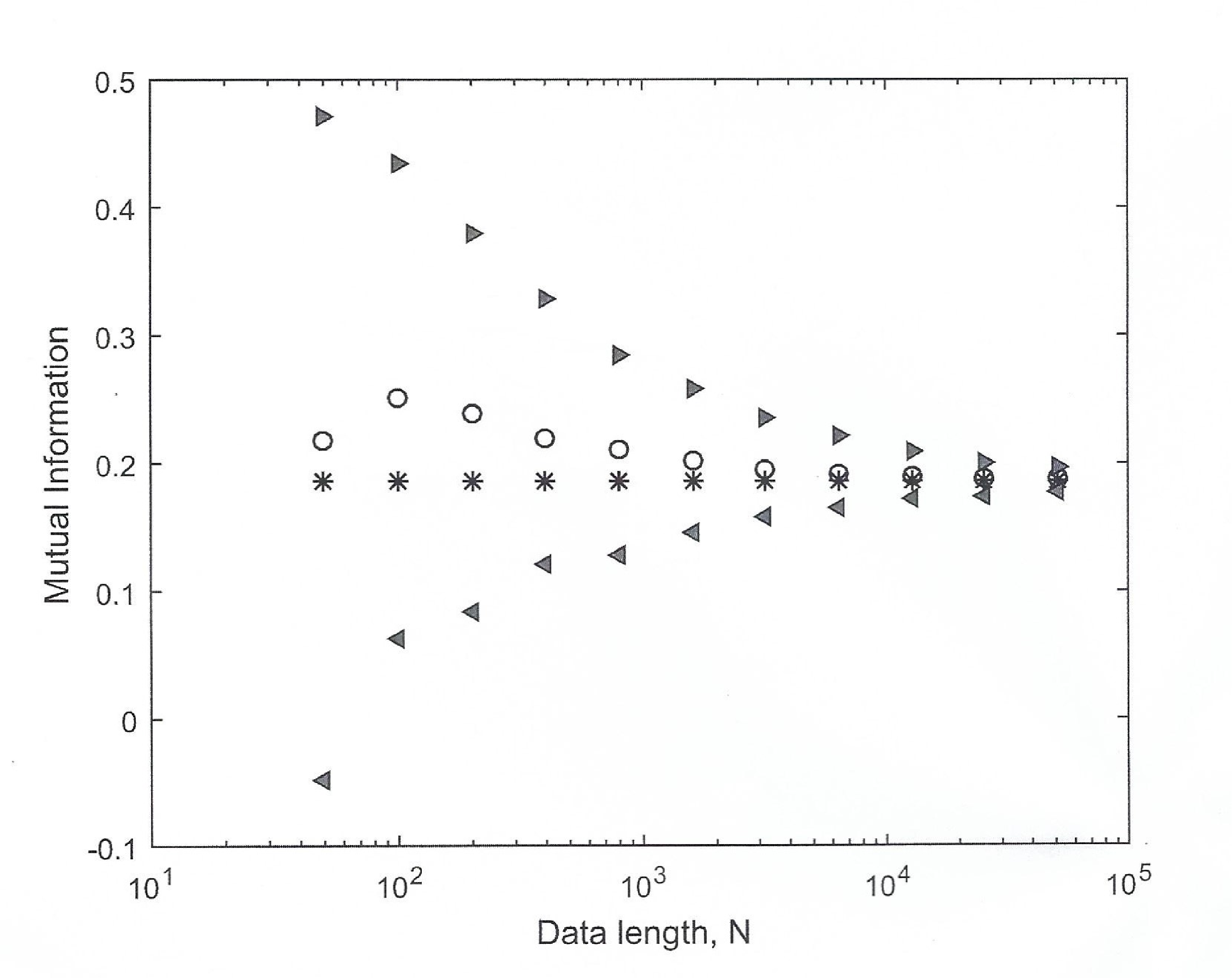}
\caption { $5\%$ and $95\%$ quantile confidence intervals ($`` \bigtriangleup $"), the estimated mean MI value (``o"), and 
the analytical solution (``*") as functions of the data length $N$. This is the case of student-t distributed bivariate data,
with the degrees of freedom $\nu=3$, $\sigma_1=\sigma_2=1$, $\rho=0.5 $. The numerical method used is the KSG estimator.} 
\label{fig:image2} 
\end{figure}

\begin{figure} [p] 
\centering 
\includegraphics [width=0.8\textwidth] {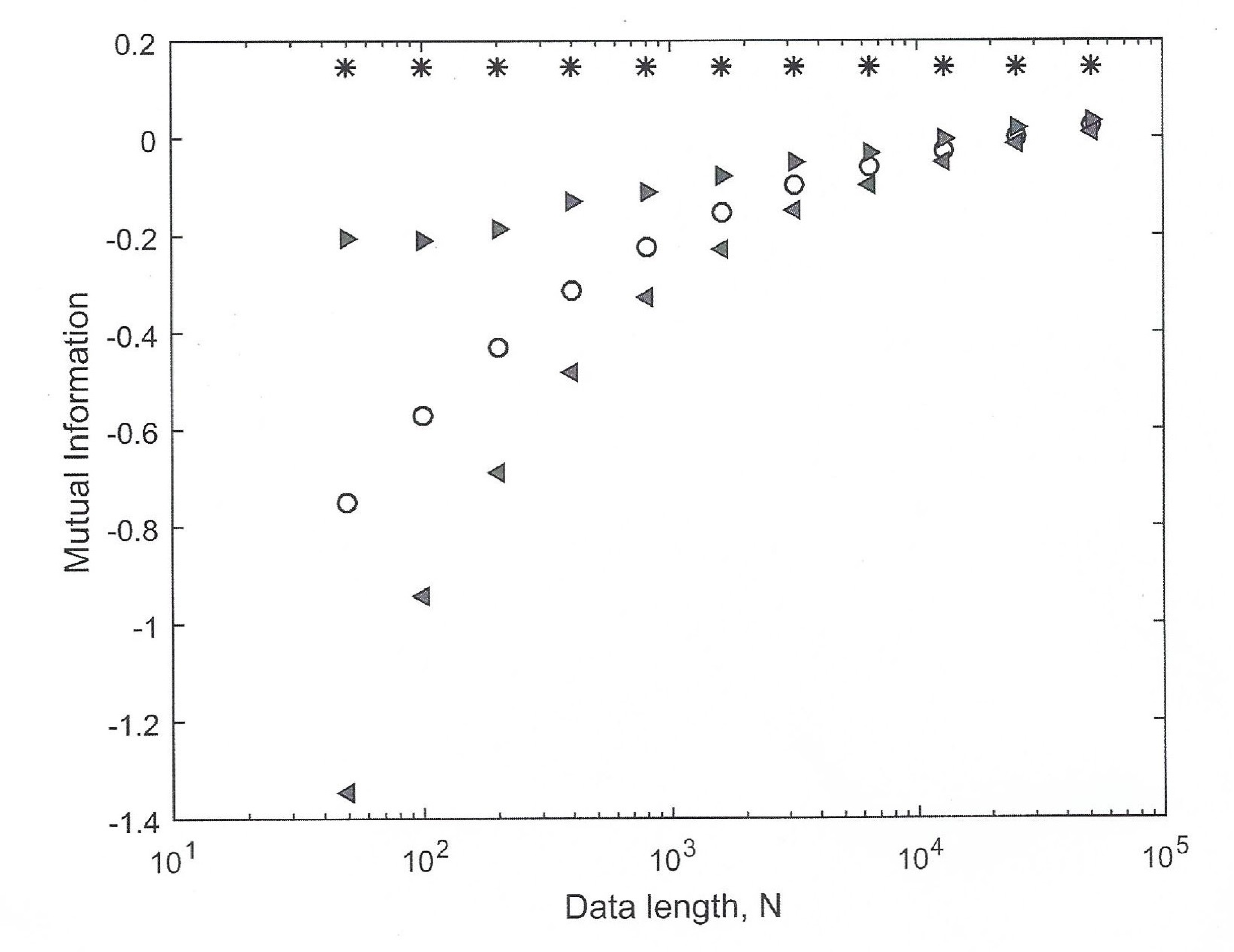} 
\caption { $5\%$ and $95 \%$ quantile confidence intervals ($`` \bigtriangleup $"), the estimated mean MI value (
``o"), and the analytical solution (``*") as functions of the data length $N$. This is the case of cubic transformed $(X', Y') = (X^3, Y^3)$ 
normal bivariate data, $\sigma_1=\sigma_2=1$, $\rho=0.5$. The numerical method used is the KSG estimator.}
\label{fig:image3} 
\end{figure}

\begin{figure} [p]
\centering 
\includegraphics [width=0.8\textwidth] {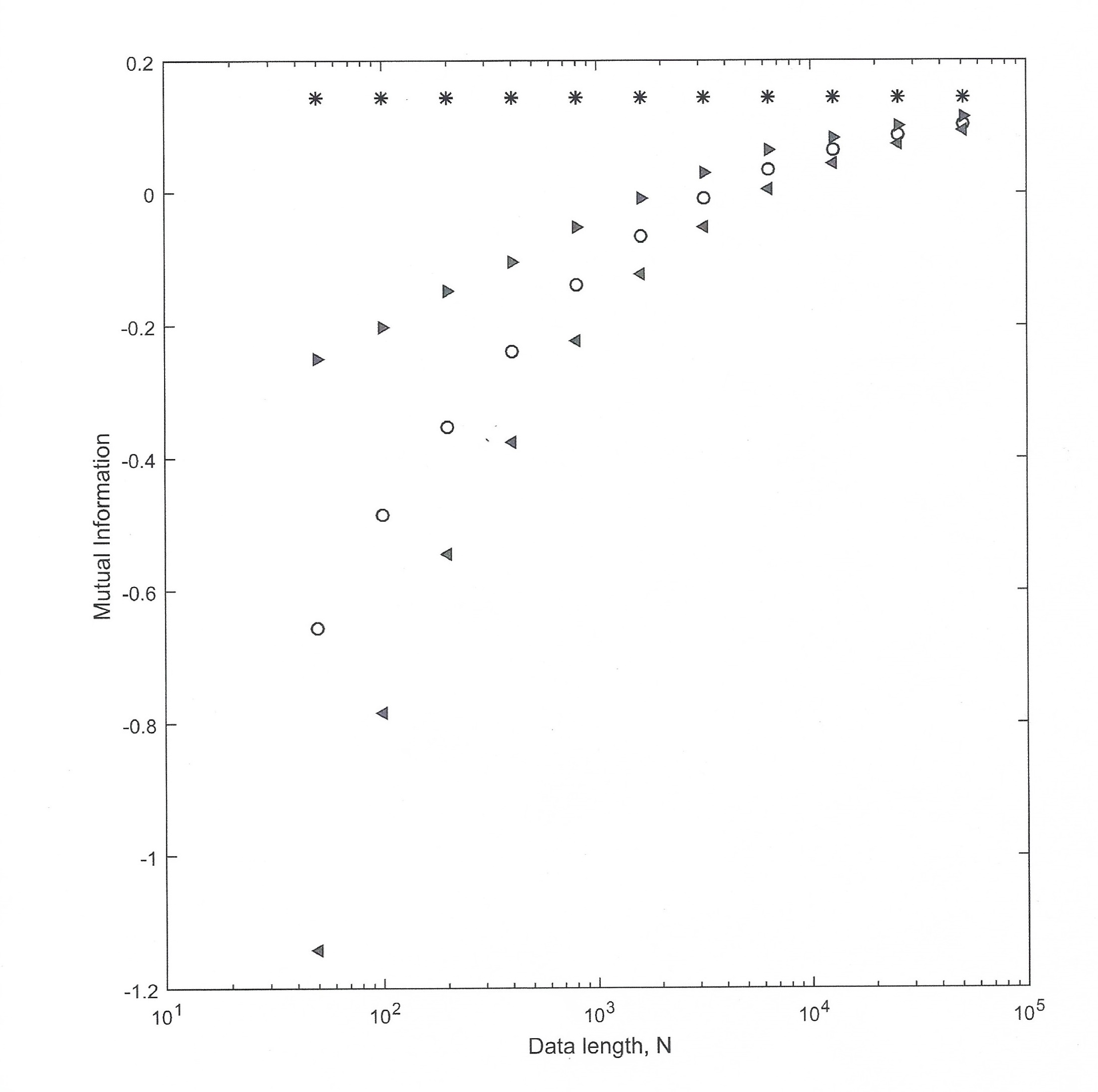} 
\caption { $5\%$ and $95\%$ quantile confidence intervals ($`` \bigtriangleup $"), the estimated mean MI value (``o"), and the
analytical solution (``*") as functions of the data length $N$. This is the case of cubic transformed $(X', Y') = (X^3, Y^3)$ lognormal 
bivariate data, $\sigma_1=\sigma_2=1$, $\rho=0.5$. The numerical method used is the KSG estimator.} 
\label{fig:image4} 
\end{figure}

\begin{figure} [p] 
\centering 
\includegraphics [width=0.8\textwidth] {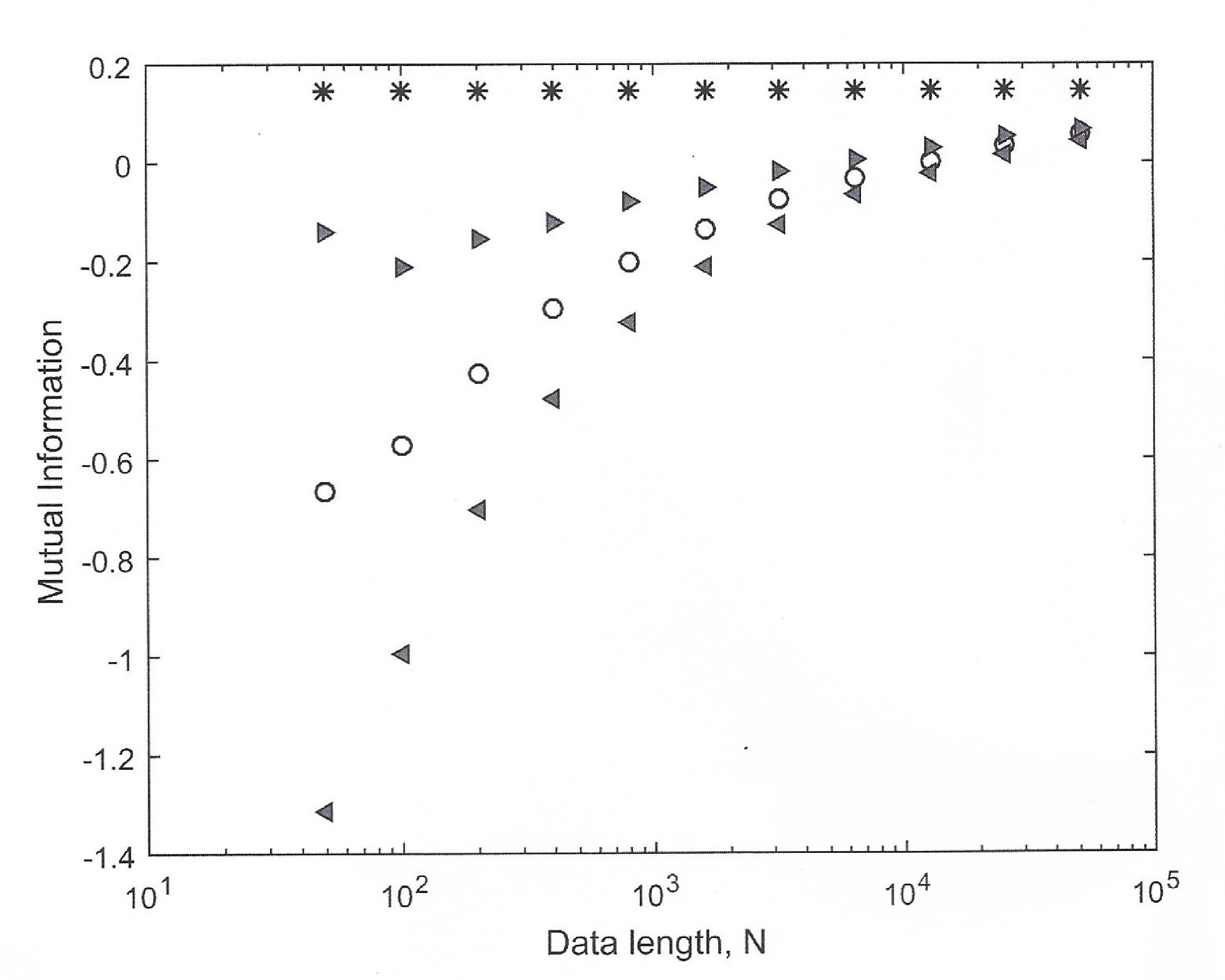} 
\caption{$5\%$ and $95\%$ quantile confidence intervals ($`` \bigtriangleup $"), the estimated mean MI value (``o"), 
and the analytical solution (``*") as functions of the data length $N$. This is the case of cubic transformed $(X', Y') = (X^3, Y^3)$ student-t 
bivariate data, with the degrees of freedom $ \nu=3$, $\sigma_1=\sigma_2=1$, $\rho=0.5$. The numerical method used is the KSG estimator.} 
\label{fig:image5} 
\end{figure}

Next, we perform calculations for Student-t bivariate data. For the simulations, we set the degrees of freedom to $\nu=3$, 
corresponding to a heavy tail for the probability density distribution function $p (x) \propto x^{-4} $. 
This distribution has a finite variance, but undefined skewness and infinite kurtosis. The covariance matrix $\in \mathbb{R}^{2 x 2}$ used 
in the Student-t bivariate distribution function is set to: $A=\begin{bmatrix} 1&0.5\\ 0.5 & 1 \end{bmatrix}$. In Figure~ \ref{fig:image2} 
resulting calculations are shown. The analytical value for the mutual information measure is given by Eq.~(\ref{studentt}). Overall, 
the performance of the KSG estimator is similar to those on the normal and lognormal bivariate data. To study the effects of a strong
non-normality on the performance of the KSG estimator, in Figures ~\ref {fig:image3},~\ref{fig:image4}, and ~\ref{fig:image5}, we present the results for the case of non-linear transformation for the normal, lognormal, and Student-t bivariate variables using $X'=X^3$, $Y'=Y^3$ rule. From these plots, one can conclude that in all cases the KSG estimator shows significant bias and fails to converge to the analytical value even for $N=10^5$. The bias does not disappear asymptotically and the confidence intervals similarly converge onto the biased estimate, leaving out the correct analytical value. This may lead to estimation errors for any confidence intervals for $MI$ which is a function of $1/N$ but does not, on average, correct for the bias in point estimates. 

Using a smaller number of runs $R = 10$, we estimate that even for the data length $N=2 \times10^6$ the bias of the estimator is about $4\%$ in the case of cubic transformation of the lognormal data, and even worse in the case of the transformed Student-t data. 
It is worth noting that in the case of $X'=X^{1/3}$, $Y'=Y^{1/3}$ transformations, the performance of the KSG estimator is significantly better.

\section {Application \& Example} 
\begin{figure} [p] 
\centering 
\includegraphics [width=0.8\textwidth] {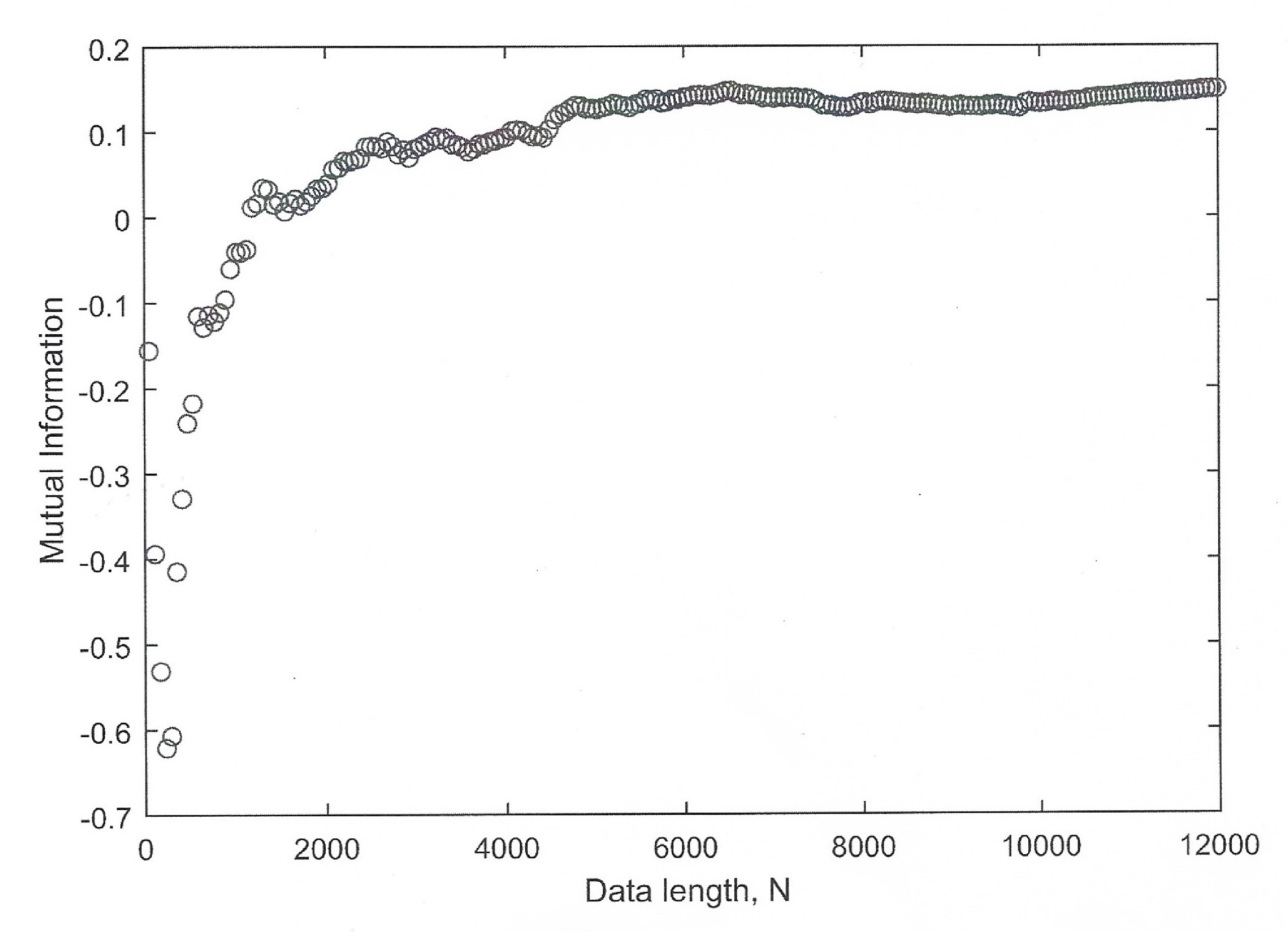} 
\caption{The estimated MI value between log returns of the CocaCola and McDonald's stocks as a function of the data length $N$.
The empirical correlation coefficient between the stock's log returns is 
$\rho=0.3975$, that would correspond to $MI=0.1241$ in a case of bivariate normal variables.
The data are boot-strapped and MI shows a smoother convergence behavior. The numerical method used is the KSG estimator.} 
\label{fig:image7} 
\end{figure}

\begin{figure} [p] 
\centering 
\includegraphics [width=0.8 \textwidth] {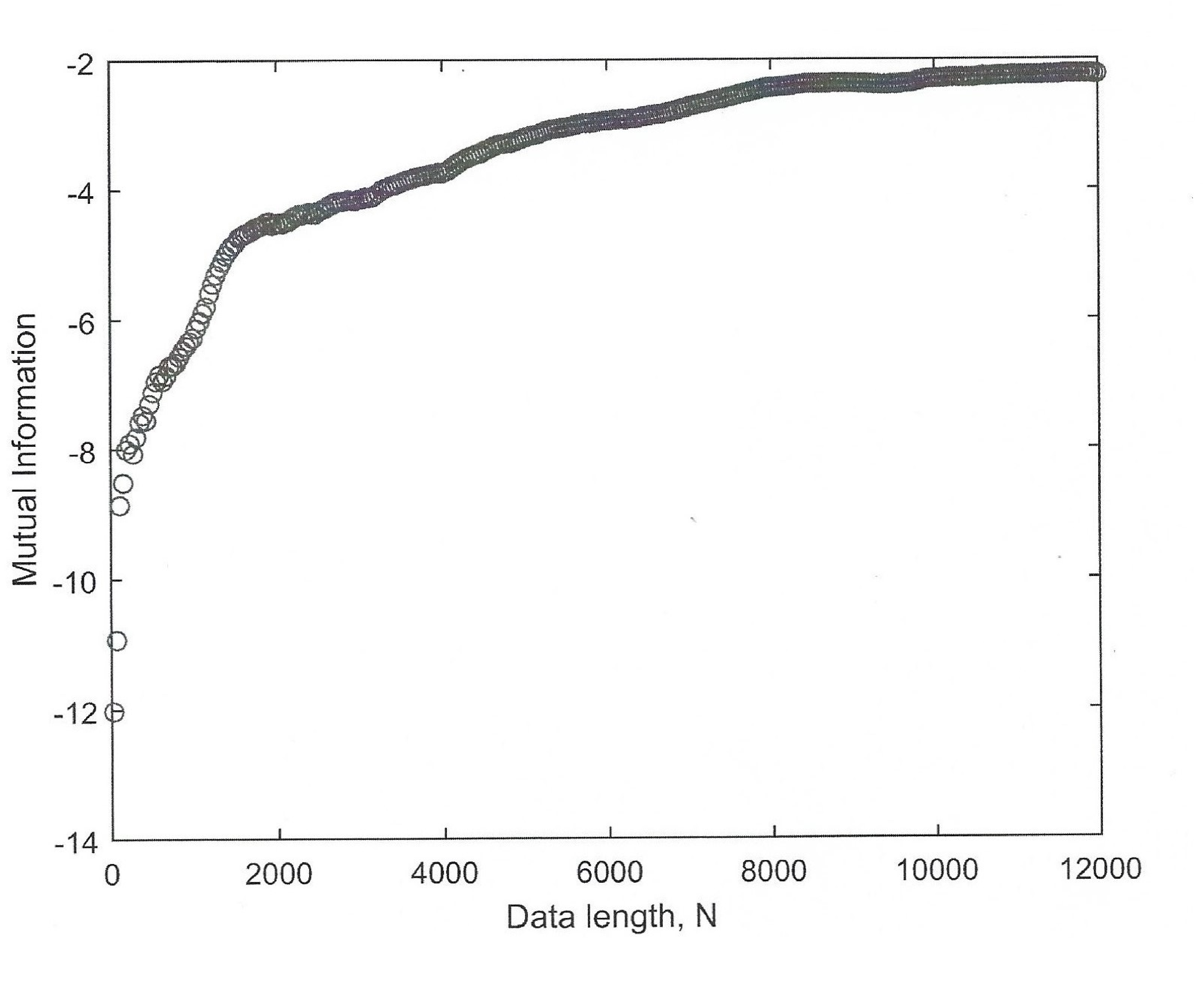} 
\caption{The estimated MI value between the cubed log returns of Coca-Cola stock and the log returns of McDonald's stock 
as a function of the data length $N$. The estimator converges to a nonsensical estimate ($MI < 0$),
demonstrating a strong bias in the estimation of  $MI$ between the two variables. The numerical method used is the KSG estimator.} 
\label{fig:image8} 
\end{figure}

Many phenomena in economics, finance, and other areas follow an empirical 
power law such as income, wealth, size of cities, and much more \cite{gaba}. In finance,
many studies find that the stock returns distribution  has
characteristics of non-normal generating processes, especially resembling those more fat-tailed than 
normal distributions or exhibiting some power law behavior \cite{fama} \cite {teich} \cite{officer}. 
For example, cross-correlation analysis between volume change and price change in financial markets \cite{podobnik} may help to understand their internal structure 
and dynamics. 
As such, calculating the empirical mutual information between 
some characteristic (e.g. prices and volume changes, etc.) of two strongly non-normal 
variables can run into misleading estimations. This can especially be a problem when making comparisons
or performing analyses on second (i.e. variance) and higher moments (e.g. skewness, kurtosis). As a practical 
example, we consider daily price data for the Coca-Cola and McDonald's stocks \cite{yahoo} 
from 1/2/1970 to 11/8/2017. The empirical correlation coefficient between the stock's log returns 
is $\rho=0.3975$, corresponding to an analytic solution of $MI\approx0.1241$ if the variables were bivariate normal;
as can be seen from Figure~\ref{fig:image7}, the empirical MI is slightly higher than the analytical estimate of 
$MI\approx0.1241$ due to the variables' non-normality. In Figure~\ref{fig:image7} we plot the estimated MI value between the 
log returns of CocaCola and McDonald's stocks as a function of the data length. We used a simple pair bootstrapping of the data and averaging over 
$50$ samples, which results in a smoother convergence behavior in Figure~\ref{fig:image7}. Finally, in Figure~\ref{fig:image8},
we compute the $MI$ between  the transformed log returns of Coca-Cola stock using the cubic nonlinearity,  and the log returns of McDonald's stock. The result displays 
a severe bias problem, ultimately converging to a $MI$ estimate of less than 0.

\section{Conclusion} We performed numerical estimations of the mutual information measure and their corresponding confidence intervals (CI). The presented results allow us to draw several conclusions. First, in the case of relatively short data length $N<300$, the Kraskov-Stoegbauer-Grassberger estimator is not reliable, especially in cases of strongly non-normal data. Our results suggest that the estimator can be considerably improved through a simple bootstrapping of the data. The simple plugin estimator has even poorer performance, when compared to the KSG approach, generating very large CIS, and is unreliable in the case of non-normal data even with a very large sample length $N\approx10^6$. In the case of large data samples $N>10^5$ one still should carefully estimate the mutual information measure confidence intervals, in particular, for data with a power-law scaling, since, in this case, 
the Kraskov-Stoegbauer-Grassberger estimator may produce significant bias. These performance issues and biases are present in both empirical data as well as simulated data. In this case, one possible way to estimate the bias is to plot the MI estimates for different data lengths on a scale $1/N$ and extrapolate in the limit as $N\to \infty$ \cite{scaling} .


\end{document}